\documentclass[aps,pra,floatfix,showpacs,noshowkeys,epsfig,graphics,natbib]{revtex4}
\usepackage{graphicx}
\usepackage{amsmath}
\usepackage{amsfonts}
\usepackage{amssymb}

\newcommand{\newc}{\newcommand}
\newc{\beq}    {\begin{equation}}
\newc{\eeq}    {\end{equation}}

\newc{\beqa}    {\begin{eqnarray}}
\newc{\eeqa}    {\end{eqnarray}}
\newc{\bs}    {\section}
\newc{\no}    {\\ \nonumber}

\newc{\st}    {\stackrel}

\begin{document}
\title{ Quantum fields as deep learning}
\author{Jae-Weon Lee}\email{scikid@jwu.ac.kr}
\affiliation{ Department of energy resources development,
Jungwon
 University,  5 dongburi, Goesan-eup, Goesan-gun Chungbuk Korea
367-805}

\date{\today}

\begin{abstract}
In this essay we conjecture that  quantum fields such as the Higgs field  is related to a restricted Boltzmann machine
for deep neural networks. An accelerating Rindler observer in a flat spacetime
sees  the quantum fields having a thermal distribution from the quantum entanglement, and
a renormalization group process for the thermal fields on a lattice is similar to a deep learning algorithm.
 This correspondence can be generalized for the KMS states of  quantum fields in a curved spacetime like a black hole.
 \end{abstract}

%\pacs{03.65.Ta,03.67.-a,04.50.Kd}
\maketitle
%\section{Introduction}

\section{Introduction}
Recently, there is a growing interest in  deep learning technology  in high energy physics, in the hope that
deep learning tools can provide  significant boost in finding  new particles at accelerators~\cite{Baldi:2014kfa}.
Deep neural networks (DNN) and the restricted Boltzmann machine (RBM) ~\cite{Hinton504}
 show unprecedent power in pattern recognitions and unsupervised learning with complex big data.
 However, the reason why deep learning can outperform other machine learning techniques in extracting features  is still unclear.
 One physical explanation is based on the analogy between the renormalization group (RG) and RBM \cite{2014arXiv1410.3831M}.
  According to the explanation RBM can mimic the coarse-graining process of RG for a thermal system and this gives the 
  efficient main feature extraction. 

Linking information science to physics is a big trend in physics nowadays. 
For example, quantum entanglement is suggested to be a source of dark energy~\cite{myDE}, gravity~\cite{Lee:2010bg,VanRaamsdonk:2016exw} and the spacetime itself~\cite{VanRaamsdonk:2010pw}.
Interestingly the holographic principle~\cite{hooft-1993} as the AdS/CFT correspondence ~\cite{Maldacena} can be also related to entanglement~\cite{Ryu:2006bv}
and RBM~\cite{Gan:2017nyt}. Motivated by these works,
in this paper we suggest that quantum field theory (QFT) can be interpreted to be a RBM and DNN.
 In Sec. II we review the relation between RG and RBM. In Sec. III an analogy between QFT  and RBM is proposed.
Section IV contains discussions.

\section{Renormalization group and Restricted Boltzmann machine}
Let us briefly review the equivalence between RG and RBM \cite{2014arXiv1410.3831M} of deep learning using
 $N$ binary spins $\mathbf v=\{v_i\}$ ($i=1,2 \cdots N$) in the Boltzmann distribution
\begin{equation}
P(\mathbf v)=\frac{e^{-\mathbf H(\mathbf v)}}{\mathcal Z},
\end{equation}
with the Hamiltonian
\begin{equation}
\label{H0}
\mathbf H(\mathbf v)=\sum_i K_i v_i +\sum_{ij} K_{ij} v_i v_j + \sum_{ijk} K_{ijk} v_i v_j v_k +\cdots,
\end{equation}
where $K_{ijk\cdots}$ are coupling constants.
Then, the partition function  $\mathcal Z$ is
\begin{equation}
\mathcal Z=\mathrm Tr_{v_i}e^{-\mathbf H(\mathbf v)},
\end{equation}
which leads to the free energy $F=-ln \mathcal Z$.
 After one step of
renormalization  one can get the effective Hamiltonian for coarse-grained block spins $\mathbf h=\{h_j\}$
\begin{equation}
\mathbf H^{RG}(\mathbf h)=\sum_i \bar K_i h_i +\sum_{ij} \bar K_{ij} h_i h_j + \sum_{ijk} \bar K_{ijk} h_i h_j h_k +\cdots,
\end{equation}
where  $\bar K_{ijk\cdots}$ are renormalized coupling constants.
 %This complete one step of renormalization.
 Repeating the above process yields
 renormalization of the theory.

In the variational RG scheme one step of RG process is implemented by introducing a function $\mathbf T_\lambda$  with some parameter $\lambda$ which satisfies
\begin{equation}
\label{T}
e^{-\mathbf H_\lambda^{RG}(\mathbf h)}\equiv \mathrm Tr_{v_i}e^{\mathbf T_\lambda (\mathbf v,\mathbf h)-\mathbf H(\mathbf v)},
\end{equation}
and then integrating out $\mathbf v$.
Here, the free energy for the coarse grained system
\beq
F_\lambda\equiv -ln(Tr_{h_j}e^{-\mathbf H_\lambda^{RG}(\mathbf h)})
\eeq
remains equal to $F$  for an exact RG process.
To do this $\mathbf T_\lambda $ should have an appropriate form.

On the other hand, Boltzmann machines are stochastic neural networks which can generate specific distribution
of data.
The restricted Boltzmann machine (RBM) is a version
composed of visible units $\mathbf v$ and hidden units $\mathbf h$ having
the following energy function describing the interaction between
the visible and  the hidden units,
\begin{align}
\label{Evh}
\mathbf E(\mathbf v,\mathbf h)=\sum_{i} b_i v_i +\sum_{j} c_j h_j +\sum_{ij}w_{ij} v_i  h_j,
\end{align}
where the units in the same layer has  no interaction between them, and
 $\lambda\equiv\{b_i,c_j,w_{ij}\}$ are variational parameters.
 The probability of a configuration of both units is given by
\begin{equation}
p_\lambda(\mathbf v, \mathbf h)= \frac{e^{-\mathbf E(\mathbf v,\mathbf h)}}{\mathcal  Z},
\end{equation}
and
 that of hidden units  by
\begin{equation}
\label{Hrbm}
p_\lambda(\mathbf h)= \sum_{\mathbf v}\frac{e^{-\mathbf E(\mathbf v,\mathbf h)}}{ \mathcal Z}
\equiv \frac{e^{-\mathbf H^{RBM}_{\lambda}(\mathbf h)}}{\mathcal  Z},
\end{equation}
which leads to the definition of
the  Hamiltonian for the hidden units $\mathbf H^{RBM}$.

An exact  mapping between the variational RG and RBM can be achieved by choosing the following function ~\cite{2014arXiv1410.3831M}
\beq
\mathbf T_\lambda (\mathbf v,\mathbf h)=-\mathbf E(\mathbf v,\mathbf h)+\mathbf H(\mathbf v).
\eeq
Then, inserting this into Eq. (\ref{T}) one can find from Eq. (\ref{Hrbm})
\beq
\mathbf H^{RG}_{\lambda}(\mathbf h)=\mathbf H^{RBM}_{\lambda}(\mathbf h),
\eeq
and similarly $\mathbf H^{RG}_{\lambda}(\mathbf v)=\mathbf H^{RBM}_{\lambda}(\mathbf v)$.
This implies that one step of the variational RG with the spins $\mathbf v$  and
$\mathbf h$
can be mapped to two layers made of units $\mathbf v$  and
$\mathbf h$  of  the RBM.

\section{Quantum field as neural networks}
How can we relate RBM with quantum fields?
Quantum fields have complex wavefunctional, hence usually do not have the Boltzmann distribution.
But, if there is a causal horizon the fields can be thermal.
For example, it is possible for an accelerating observer to see the flat spacetime vacuum state as a Boltzmann distribution,
 which is the Unruh effect.

Consider an observer with acceleration $a$ in $x_1$ direction
 with coordinates $(t,x_1,x_2,x_3)$  in a flat spacetime, who observes
 a scalar field with Hamiltonian
 \beq
 H(\phi)=\int d^3 x \left[ \frac{1}{2}\left( \frac{\partial \phi}{\partial t}\right)^2+
 \frac{1}{2}\left( {\nabla \phi}\right)^2 + V(\phi) \right]
 \eeq
 with potential $V$. The field could be the standard model Higgs, inflaton or ultra-light scalar dark matter~\cite{Lee:2017qve}.
 The Rindler coordinates $(\eta,r,x_2,x_3)$ can be defined with
\beq
\label{rindler}
t= r~ sinh (a \eta),~ x_1= r ~cosh (a \eta)
\eeq
 on the Rindler wedges.

 In the Rindler coordinates  the proper time interval is
  $ard\eta$ and hence the corresponding Hamiltonian becomes
  \beqa
 H_R =  \int_{} dr dx_\bot~ ar
 \left[
  \frac{1}{2}\left( \frac{\partial \phi}{ar \partial \eta}\right)^2
 +
 %\no
 \frac{1}{2}
 \left(\frac{\partial \phi}{ \partial r} \right)^2+\frac{1}{2}
 \left( {\nabla_\bot \phi}\right)^2
 +  V(\phi)\right] ,
 \eeqa
where $\bot$ denotes the spatial direction orthogonal to $(\eta,r)$.
Then, the Rinder observer sees a horizon at $r=0$.

It is well-known that
we can decompose the fields in the  left and right Rindler wedges as $\phi_L$ and $\phi_R$, respectively, and
the ground state of $H_R$ is then described by a
 wavefunctional
\beq
 \Psi_0(\phi_L, \phi_R)=\frac{1}{\sqrt{\mathcal Z}}\langle \phi_L|e^{-\pi H_R}|\phi_R\rangle.
\eeq
The two fields are entangled, and
the reduced density matrix for $\phi_R$ is given by partial tracing $\phi_L$, i.e.,
$\rho_R=Tr_{\phi_L} \Psi_0 \Psi_0^\dagger=   \frac{1}{Z}exp(-2\pi H_R)$. With the proper redshifted Unruh  temperature $T=a/2\pi$
this density matrix becomes
\beq
\rho_R=\frac{1}{\mathcal Z}exp(- H_R/T),
\eeq
which means $\phi_R$ has a Boltzmann distribution, and
the Minkowski vacuum restricted to the one Rindler wedge is a KMS state ~\cite{Ross:2005sc}.

Now, we suggest that
the quantum fields $\phi_R$ can be treated as a continuous version of $\mathbf v$, and $H_R$ can be $\mathbf H(\mathbf v)$ in Eq. (\ref{H0}) for RBM.
Recall that the RG process  is a natural process in QFT.
We propose that  the coarse graining process for the quantum field corresponds to the information propagation in the deep neural networks.
To be specific, let us consider a discretized spacetime with the minimum length scale $l$ of order of the Planck scale
as in the lattice field theory. We also assume a quadratic potential with mass $m$. Then, in $d+1$ spacetime with
a  field $\phi_R$ at the site $x$, $\phi_{x}$,
\beq
 H_R \simeq   N_1 l^{d+1}
\sum_{r}a r \sum_{x}\left[\frac{(\phi_{x+\eta}-\phi_x )^2}{2(ar l)^2}
+ \sum_{\mu=1}^d\frac{(\phi_{x+\hat{\mu}}-\phi_x )^2}{2l^2} +\frac{m^2 \phi_x^2}{2} \right],
\eeq
where $N_1$ is a normalization, $\hat{\mu}$ represents the unit vectors to the nearest  points in the spatial direction $\mu$,
and $\{x,r,\eta\}$ should be understood to be integer indexes ($r\ge 1$).
With an appropriate $N_1$ we can rescale the field as  $0\le \phi_x \le 1$. This can be justified because physical $\phi_x$ can not have
an arbitrary large value, and hence there should be a maximum field value, say, of order of the Planck mass.

Now, with $H_R$ and $E(\mathbf v,\mathbf h)$ in Eq. (\ref{Evh}) we can perform the one step of variational RG using
Eq. (\ref{T}). Here, the lattice field $\phi_x$ plays a role of the visible unit  $v_i$ and
renormalized field $\tilde{\phi}_x$ plays a role of the hidden unit  $h_j$.
At the next level  $\tilde{\phi}_x$ acts as a new visible unit, and one can repeat the RG steps toward the IR limit.
Therefore, the RG process for the scalar field corresponds to DNN and it is a kind of natural learning process. (See Fig. 1)
\begin{figure}[tpbh]
\includegraphics[width=0.6\textwidth]{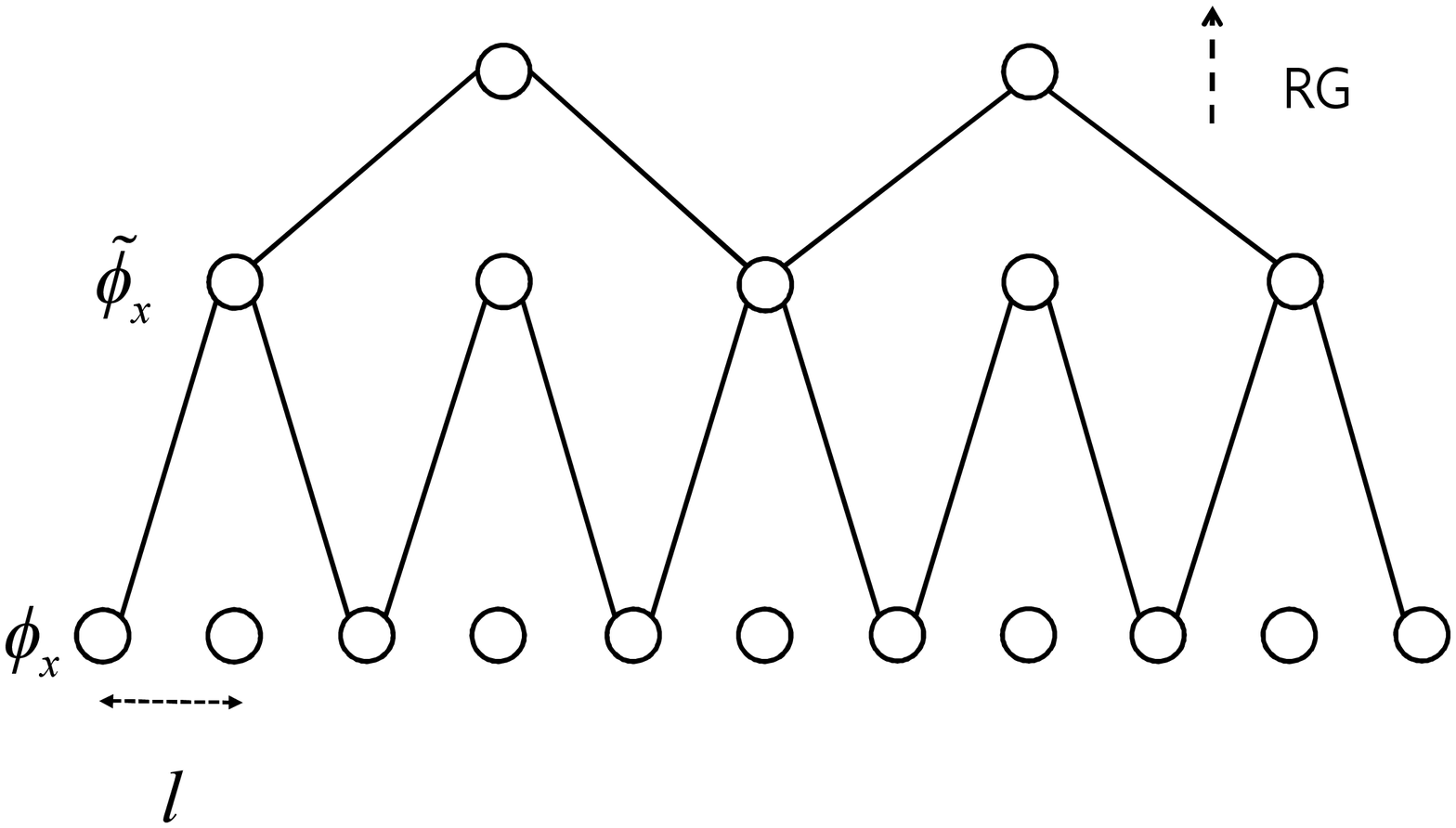}
\caption{ Quantum fields $\phi_x$ on a lattice with a UV-cutoff $l$ act as visible units.
The solid lines represent $w_{ij}$.
After one step of the variational RG with the decimation, the renormalized field  $\tilde{\phi}_x$  plays a role of the hidden units.
At the next level $\tilde{\phi}_x$ acts as a new visible units.
The whole RG process then corresponds to DNN.}
\end{figure}

At each RG step, there is a coarsegraining of the field  leading to effective field theory of the system.
Like the output units in RBM, this effective field contain the concise information of the lower units, that is, UV-physics.
This might explain why effective field theory is so successful to describe a low energy physics despite of partial information loss
about the UV-phyics.
Repeating the real space RG steps leads to the RG process toward an IR region, which corresponds to DNN.
One can check the validity of this concept by reverting the process and
 approximately reproducing the input information (field values of the lowest units in the Fig. 1) from the output units (the most upper units) and the trained parameters  $\{b_i,c_j,w_{ij}\}$.
This corresponds to  the inversion of the ordinary RG process in the field theory from IR to UV.

Further simplification can be done for a numerical study by considering an Rindler observer with a hugh acceleration $a\gg 1$. Then,
we can ignore the time derivative term and get
\beq
 H_R \simeq   N_1
\sum_{r}\frac{a r}{2} \sum_{x}\left[\sum_{\mu=1}^d{(\phi_{x+\hat{\mu}}-\phi_x )^2} +{m^2 \phi_x^2} \right],
\eeq
where we set $l=1$. From the above equation we expect the thermal fluctuation of the field mainly exists near the horizon, i.e., $r\simeq 1$.

We have considered the vacuum state so far. For a slightly excited state $\Psi_0+\delta \Psi$, the initial density matrix and  the probability distribution
should be slightly changed. This effect can be reflected by including
 an  interaction  term  $H_{int}$ into the Hamiltonian $H_R$ . Otherwise, if we keep $H_R$ fixed,  $ E(\mathbf v,\mathbf h)$ and
the couplings $\{b_i,c_j,w_{ij}\}$ should be changed instead to represents the excited state. This might be
another kind of natural learning process. Thus, we guess there is a mapping between quantum states not far from the vacuum state and information (i.e., parameters) in the corresponding RBM model.

It is straightforward to extend the previous arguments to a black hole case.
For the Schwarzchild black holes with mass $M$  the metric is given by
\beq
ds^2=-Fdt^2+F^{-1}dr^2+r^2d\Omega^2,
\eeq
where $F=1-2GM/r$.
Near the event horizon this reduces to the Rindler metric
\beq
ds^2\simeq -R^2 d\eta^2+dR^2+r^2 d\Omega^2,
\eeq
with $R=\sqrt{r(r-2MG)}$ and $\eta=t/4GM$ as is well-known.
Therefore, we expect quantum fields near the black hole horizon is also a KMS state and can be  viewed as a DNN for a observer
seeing the Hawking radiation.

\section{Discussions}

Yet another possible approach is to use the well-known correspondence of the Euclidean quantum field theory in $d+1$ dimensional flat spacetime and the statistical mechanics in $d+1$ dimensional flat space
using  an imaginary time.  In this case we do not need an accelerating observer.
The Euclidean functional integral
\beq
\mathcal Z=\int d\phi e^{-\int d^{d+1}x~ H(\phi)/\hbar}
\eeq
has the form of the partition function for the classical thermal system with $T=\hbar$
and one can now easily see the analogy to DNN.

It would be easy to extend our arguments to the KMS states of other spin fields such as fermions, gauge vectors, and gravitons
with causal horizons.
The unexpected relation between the quantum field and DNN
 might explain why DNN is so successful in particle identification at accelerator experiments ~\cite{Baldi:2014kfa}.
 Conversely, QFT can give some insights to understand why RBM is so powerful.

Our conjecture also implies  a surprising possibility that
the quantum fields, and hence matter in the universe, can memorize information and even can perform self-learning to some extend
like DNN in a way consistent with the Strong Church-Turing thesis.

%\section*{acknowledgments}
\acknowledgments
This work was supported by the Jungwon University Research Grant (2016-040).

%\bibliographystyle{h-physrev}
%\bibliography{entanglement}

\end{document}